\documentclass[aps,prb,twocolumn,showpacs]{revtex4}

\usepackage{graphicx}
\usepackage{amsmath}
\usepackage{amssymb}

\begin{document}

\title{Hidden pair-density-wave order in cuprate superconductors}

\author{Shiping Feng\footnote{Electronic address: spfeng@bnu.edu.cn}, Deheng Gao, Yiqun Liu, Yingping Mou, and Shuning Tan}


\affiliation{Department of Physics, Beijing Normal University, Beijing 100875, China}

\begin{abstract}
When the Mott insulating state is suppressed by charge carrier doping, the pseudogap phenomenon emerges, where at the low-temperature limit, superconductivity coexists with some ordered electronic states. Within the framework of the kinetic-energy-driven superconductivity, the nature of the pair-density-wave order in cuprate superconductors is studied by taking into account the pseudogap effect. It is shown that the onset of the pair-density-wave order does not produce an ordered gap, but rather a novel hidden order as a result of the interplay of the charge-density-wave order with superconductivity. As a consequence, this novel hidden pair-density-wave order as a subsidiary order parameter coexists with the charge-density-wave order in the superconducting-state, and is absent from the normal-state.
\end{abstract}


\maketitle

One of the most remarkable features in cuprate superconductors is the presence of the pseudogap in the temperature range above the superconducting (SC) transition temperature $T_{\rm c}$ but below the pseudogap crossover temperature $T^{*}$ \cite{Hufner08,Basov05,Timusk99}. In particular, the experimental results from the angle-resolved photoemission spectroscopy (ARPES) observations indicated that this pseudogap can be identified as being a region of the self-energy effect in which it suppresses the spectral weight of the low-energy quasiparticle excitation spectrum \cite{Norman98,Yoshida06,Kanigel07,Yoshida09,Yang08,Meng09,Yang11}, where the low-energy spectral weight of the quasiparticle excitation spectrum on the electron Fermi surface (EFS) around the antinodal region is gapped out by the pseudogap, leaving behind the low-energy spectral weight of the quasiparticle excitation spectrum only located at the disconnected segments around the nodal region to form the Fermi pockets. This Fermi pocket consists of the Fermi arc and back side of Fermi pocket. However, the experimental results observed recently from different measurement techniques  \cite{Comin16,Wu11,Chang12,Ghiringhelli12,Comin14,Neto14,Fujita14,Campi15,Comin15a,Peng16,Hinton16,Hamidian16} demonstrated that the charge-density-wave (CDW) order exists within the pseudogap phase, appearing below a temperature $T_{\rm CDW}$ well above $T_{\rm c}$ in the underdoped regime, and coexists with superconductivity below $T_{\rm c}$. $T_{\rm CDW}$ is the temperature where the CDW order develops, and is of the order of $T^{*}$. Furthermore, the CDW vector $Q_{\rm CDW}$ was reported to connect two tips of the straight Fermi arcs \cite{Comin16,Wu11,Chang12,Ghiringhelli12,Comin14,Neto14,Fujita14,Campi15,Comin15a,Peng16,Hinton16,Hamidian16}. These experimental results show that the CDW order plays a very important role in the physics of the pseodogap phase of cuprates superconductors \cite{Comin16,Wu11,Chang12,Ghiringhelli12,Comin14,Neto14,Fujita14,Campi15,Comin15a,Peng16,Hinton16,Hamidian16}. On the other hand, the measurements of the Hall coefficient in high magnetic fields indicated that the EFS reconstruction by the CDW order ends at a critical doping that is lower than the pseudogap critical point \cite{Badoux16}, which therefore show that the pseudogap and CDW order are separate phenomena \cite{Badoux16}.

Despite the general consensus about the existence of the CDW order in cuprate superconductors \cite{Comin16,Wu11,Chang12,Ghiringhelli12,Comin14,Neto14,Fujita14,Campi15,Comin15a,Peng16,Hinton16,Hamidian16,Badoux16}, its physical origin is still controversial. Theoretically, two main scenarios are disputing the explanation of the nature of the CDW order. In one of the scenarios \cite{Sachdev13,Meier13,Harrison14,Atkinson15,Feng16,Zhao16,Gao18}, the CDW vectors spanning the tips of the Fermi arcs are a manifestation of the pseudogap formation due to the CDW order, while in the other \cite{Lee14,Fradkin15,Wang15}, the pesudogap phase is attributed to the pair-density-wave (PDW) state, where electrons on the same side of EFS are paired, and therefore is different from the CDW state. In this scenario, the CDW order only appears in the pesudogap phase as a subsidiary order parameter, and the tips of the Fermi arcs themselves result from an EFS instability around the antinodal region that is distinct from the CDW order. In this case, some natural questions are raised: (i) what type ordered-state is the crucial ordered-state for the competition with superconductivity? (ii) does the PDW order coexist with the CDW order and superconductivity? In our recent studies \cite{Feng16,Zhao16,Gao18}, the physical origin of the CDW order and of its interplay with superconductivity in cuprate superconductors have been studied within the framework of the kinetic-energy-driven SC mechanism by taking into account the pseudogap effect, where the main features of the CDW order in cuprate superconductors are qualitatively reproduced \cite{Comin16,Wu11,Chang12,Ghiringhelli12,Comin14,Neto14,Fujita14,Campi15,Comin15a,Peng16,Hinton16,Hamidian16}, including the doping dependence of the CDW vector. In particular, we show that the CDW order both in the SC- and normal-states is driven by the pseudogap-induced EFS instability, with the CDW vector that is well consistent with the wave vector connecting the straight tips of the Fermi arcs. However, although the CDW order coexists with superconductivity below $T_{\rm c}$, this CDW order antagonizes superconductivity \cite{Gao18}. In this paper, we try to discuss the physical origin of the PDW order in cuprate superconductors and of its connection with the interplay of the CDW order and superconductivity along with this line. We show that the PDW order is generated by the interplay between the CDW order and superconductivity, therefore this PDW order appears automatically in the SC-state as a subsidiary order parameter, and is absent from the normal-state. Moreover, we demonstrate that the onset of the PDW order does not produce an ordered gap, and in this sense, the PDW order is a novel hidden order.

Our present study of the physical origin of the hidden PDW order in cuprate superconductors is based on the framework of the kinetic-energy-driven superconductivity \cite{Feng0306,Feng15}, which has been employed previously to study the unusual behavior of the SC-state quasiparticle excitations in cuprate superconductors \cite{Feng15a,Gao18a} and the related interplay of the CDW order and superconductivity \cite{Gao18}. Most importantly, this kinetic-energy-driven superconductivity is associated with the single-particle diagonal and off-diagonal Green's functions $G({\bf k}, \tau-\tau')=-<T_{\tau}C_{{\bf k}\sigma}(\tau)C^{\dagger}_{{\bf k}\sigma}(\tau')>$ and $\Im^{\dagger}({\bf k},\tau-\tau')=<T_{\tau} C^{\dagger}_{{\bf k}\uparrow}(\tau)C^{\dagger}_{-{\bf k}\downarrow}(\tau')>$ of the $t$-$J$ model in the charge-spin separation fermion-spin representation, which have been derived within the framework of the full charge-spin recombination as \cite{Feng15a},

\begin{widetext}
\begin{subequations}\label{EGF}
\begin{eqnarray}
G({\bf k},\omega)&=&{1\over \omega-\varepsilon_{\bf k}-\Sigma_{1}({\bf k},\omega)-[\Sigma_{2}({\bf k},\omega)]^{2}/[\omega+\varepsilon_{\bf k}+ \Sigma_{1}({\bf k},-\omega)]}, ~~~~~~ \label{DEGF}\\
\Im^{\dagger}({\bf k},\omega)&=&-{\Sigma_{2}({\bf k},\omega)\over [\omega-\varepsilon_{\bf k}-\Sigma_{1}({\bf k},\omega)][\omega+\varepsilon_{\bf k}+ \Sigma_{1}({\bf k},-\omega)]-[\Sigma_{2}({\bf k},\omega)]^{2}},~~~~\label{ODEGF}
\end{eqnarray}
\end{subequations}
\end{widetext}
\noindent
where the bare electron excitation spectrum $\varepsilon_{\bf k}=-Zt\gamma_{\bf k}+Zt'\gamma_{\bf k}'+\mu$, with $\gamma_{\bf k}=({\rm cos}k_{x}+{\rm cos}k_{y})/2$, $\gamma_{\bf k}'= {\rm cos} k_{x}{\rm cos}k_{y}$, $t$ and $t'$ are the nearest-neighbor (NN) and next NN electron hopping amplitudes in the $t$-$J$ model, respectively, $Z$ is the number of the NN or next NN sites on a square lattice, and $\mu$ is the chemical potential, while the electron self-energies $\Sigma_{1}({\bf k},\omega)$ in the particle-hole channel and $\Sigma_{2}({\bf k},\omega)$ in the particle-particle channel originated from the interaction of electrons with spin excitations have been evaluated in terms of the full charge-spin recombination, and are given explicitly in Ref. \onlinecite{Feng15a}.

In the framework of the kinetic-energy-driven superconductivity, the electron self-energy $\Sigma_{2}({\bf k},\omega)$ describes a coupling of the electron pair interaction strength and electron pair order parameter, and in the static limit, it thus is defined as the momentum dependence of the SC gap \cite{Feng15a,Gao18} $\bar{\Delta}_{\rm s}({\bf k})=\Sigma_{2}({\bf k},\omega=0)$. On the other hand, the electron self-energy $\Sigma_{1}({\bf k}, \omega)$ describes the single-particle coherence, and therefore is closely related to the pseudogap as \cite{Feng15a,Gao18},
\begin{eqnarray}\label{PG}
\Sigma_{1}({\bf k},\omega)\approx {[\bar{\Delta}_{\rm PG}({\bf k})]^{2}\over\omega+\varepsilon_{0{\bf k}}},
\end{eqnarray}
where the energy spectrum $\varepsilon_{0{\bf k}}$ and the momentum dependence of the pseudogap $\bar{\Delta}_{\rm PG}({\bf k})$ have been obtained explicitly in Ref. \onlinecite{Feng15a}.

\begin{figure*}[t!]
\centering
\includegraphics[scale=0.40]{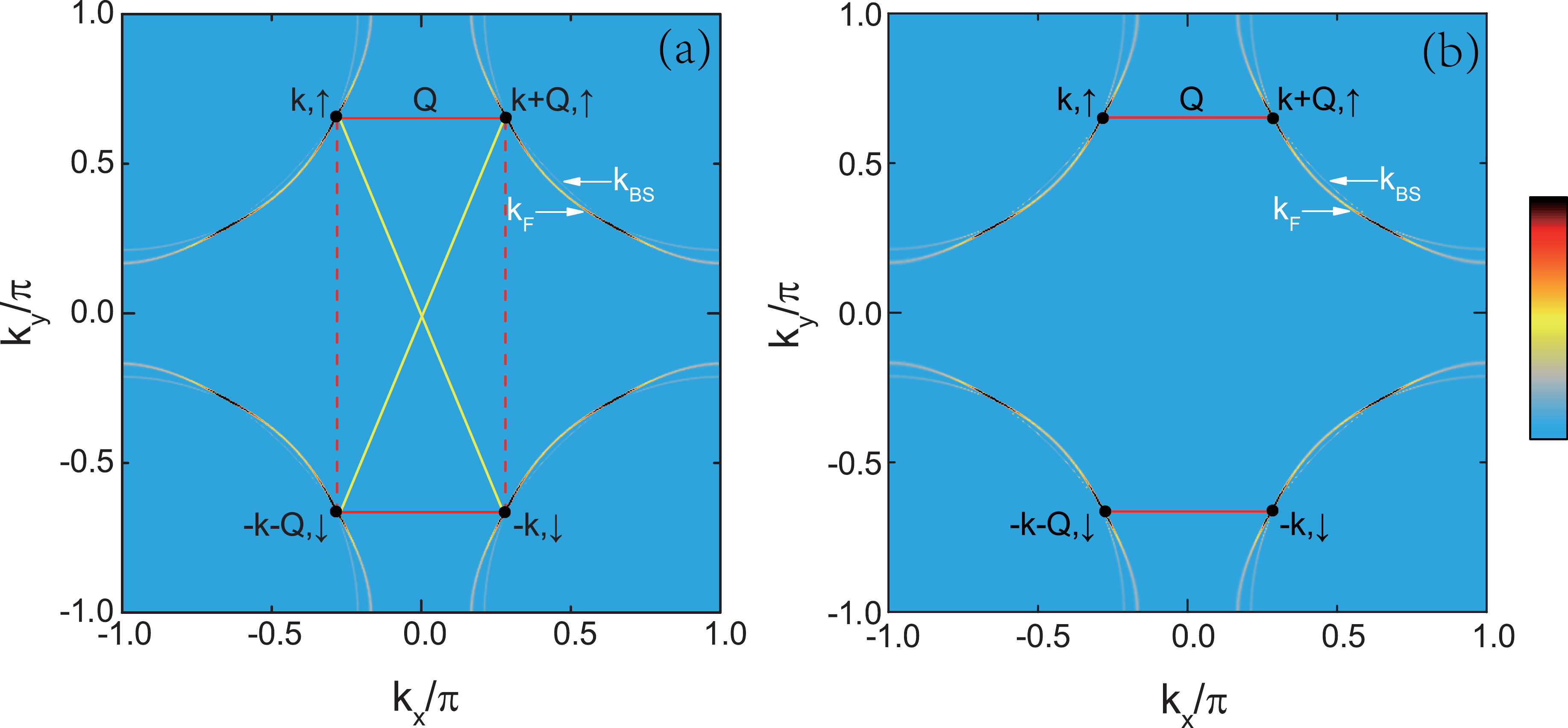}
\caption{(Color online) The intensity map of the quasiparticle excitation spectrum $I({\bf k},0)$ in the $[k_{x},k_{y}]$ plane in (a) the SC-state and (b) the normal-state at $\delta=0.09$ with $T=0.002J$ for $t/J=2.5$ and $t'/t=0.3$. The pairing of electrons and holes at ${\bf k}$ and ${\bf k} +{\bf Q}$ [red lines in (a) and (b)] drives the CDW order formation, and the electron pairing at ${\bf k}$ and ${-{\bf k}-{\bf Q}}$ states [dashed-red lines in (a)] governs the hidden PDW order, while the electron pairing at ${\bf k}$ and ${-\bf k}$ states [yellow lines in (a)] is responsible for superconductivity. \label{spectrum-maps}}
\end{figure*}

With the help of the above single-particle diagonal Green's function (\ref{DEGF}), we can obtain the quasiparticle excitation spectrum $I({\bf k}, \omega)\propto n_{\rm F}(\omega)A({\bf k},\omega)$, with the fermion distribution $n_{\rm F}(\omega)$ and the electron spectral function $A({\bf k}, \omega)=-2{\rm Im}G({\bf k},\omega)$. This quasiparticle excitation spectrum therefore describes the energy and momentum dependence of the ARPES spectrum \cite{Damascelli03,Campuzano04,Zhou07}. On the experimental hand, the combination of the resonant X-ray scattering (RXS) data and EFS measured results using ARPES have revealed a quantitative link between the CDW vector $Q_{\rm CDW}$ and the momentum vector connecting the tips of the straight Fermi arcs \cite{Comin16,Comin14,Neto14}, which in this case coincide with the hot spots on EFS. However, on the theoretical hand, we \cite{Feng16,Zhao16,Gao18} have studied the quantitative connection between the collective response of the electron density and the low-energy electronic structure within the framework of the kinetic-energy-driven superconductivity, and the obtained results are well consistent with these RXS and ARPES experimental data \cite{Comin16,Comin14,Neto14}. In Fig. \ref{spectrum-maps}, we plot the quasiparticle excitation spectrum $I({\bf k},0)$ as a function of the momentum in (a) the SC-state and (b) the  normal-state at doping $\delta=0.09$ with temperature $T=0.002J$ for parameters $t/J=2.5$ and $t'/t=0.3$. Obviously, the results in Fig. \ref{spectrum-maps} indicates that there are two continuous contours in momentum space, which are labeled as ${\bf k}_{\rm F}$ and ${\bf k}_{\rm BS}$, respectively. However, some striking features appear: (A) The low-energy spectral weight on the constant energy contours ${\bf k}_{\rm F}$ and ${\bf k}_{\rm BS}$ around the antinodal region has been gapped out by the pseudogap, and then the low-energy quasiparticle excitations occupy disconnected segments located at the contours ${\bf k}_{\rm F}$ and ${\bf k}_{\rm BS}$ around the nodal region; (B) However, the highest peak heights are located at the tips of the disconnected segments, where the most quasiparticles are accommodated; (C) The tips of the disconnected segments on the contours ${\bf k}_{\rm F}$ and ${\bf k}_{\rm BS}$ converge at the hot spots to form a closed Fermi pocket \cite{Yang08,Meng09,Yang11,Zhao17}, with the disconnected segment at the first contour ${\bf k}_{\rm F}$ is called the Fermi arc, while the other at the second contour ${\bf k}_{\rm BS}$ is associated with the back side of the Fermi pocket; (D) These Fermi pockets appear both in the SC- and normal-states, and are not symmetrically located in the Brillouin zone, i.e., they are not centered around $[\pm\pi/2,\pm\pi/2]$; (E) The quasiparticle scattering wave vector between the tips of the straight Fermi arcs both in the SC- and normal-states shown in Fig. \ref{spectrum-maps} (red lines) at the underdoping $\delta=0.09$ is $Q_{\rm HS}=0.280$ (hereafter we use the reciprocal units), which is in good agreement with the experimental average value of the CDW vector $Q_{\rm CDW}\approx 0.29$ observed both in the SC- and normal-states of the underdoped cuprate superconductors \cite{Comin16,Wu11,Chang12,Ghiringhelli12,Comin14,Neto14,Fujita14,Campi15,Comin15a,Peng16,Hinton16,Hamidian16}, indicating that the CDW order both in the SC- and normal-states is driven by the EFS instability. However, it should be emphasized that the momentum dependence of the CDW gap at the tips of the Fermi arcs is very small ($\sim 0$) \cite{Comin16,Wu11,Chang12,Ghiringhelli12,Comin14,Neto14,Fujita14,Campi15,Comin15a,Peng16,Hinton16,Hamidian16}, which directly contradicts the standard CDW picture where an energy gap is expected at precisely that points. Moreover, we have also shown that this CDW vector is doping dependent, with the magnitude of the CDW vector $Q_{\rm CDW}$ that decreases with the increase of doping, also in good agreement with experimental results \cite{Comin16,Comin14,Neto14}.

We are now ready to discuss the physical origin of the PDW order in cuprate superconductors and of its connection with the interplay between the CDW order and superconductivity. In the above discussions, the quasiparticle excitation spectrum $I({\bf k},\omega)$ is obtained in terms of the electron spectral function, and therefore the essential behavior of the quasiparticle excitations in cuprate superconductors is completely determined by the electron spectral function [then the single-particle diagonal Green's function (\ref{DEGF}) and the related electron self-energy (\ref{PG})]. However, we find that these single-particle diagonal and off-diagonal Green's functions in Eq. (\ref{EGF}) and the related electron self-energy in Eq. (\ref{PG}) can be also reproduced exactly by a phenomenological Hamiltonian,
\begin{eqnarray}\label{CDW-model}
H&=&\sum_{{\bf k}\sigma}\varepsilon_{\bf k}C^{\dagger}_{{\bf k}\sigma}C_{{\bf k}\sigma}-\sum_{{\bf k}\sigma}\varepsilon_{0{\bf k}}C^{\dagger}_{{\bf k} +{\bf Q}\sigma}C_{{\bf k}+{\bf Q}\sigma}\nonumber\\
&+&\sum_{{\bf k}\sigma}\bar{\Delta}_{\rm PG}({\bf k})(C^{\dagger}_{{\bf k}+{\bf Q}\sigma}C_{{\bf k}\sigma}+C^{\dagger}_{{\bf k}\sigma} C_{{\bf k}+ {\bf Q}\sigma})\nonumber\\
&-&\sum_{\bf k}\bar{\Delta}_{\rm s}({\bf k})(C^{\dagger}_{{\bf k}\uparrow}C^{\dagger}_{-{\bf k}\downarrow}+C_{-{\bf k}\downarrow}C_{{\bf k} \uparrow}), ~~~~~~~~~
\end{eqnarray}
where the dispersions of $\varepsilon_{\bf k}$ and $\varepsilon_{0{\bf k}}$, the SC gap $\bar{\Delta}_{\rm s}({\bf k})$, and the pseudogap  $\bar{\Delta}_{\rm PG}({\bf k})$ are given explicitly in Eqs. (\ref{EGF}) and (\ref{PG}). In particular, the pseudogap $\bar{\Delta}_{\rm PG}({\bf k} )$ has been identified as the momentum dependence of the CDW gap. In our previous discussions, we \cite{Zhao16} have shown that this pseudogap $\bar{\Delta}_{\rm PG}({\bf k})$ has a strong angular dependence, where $\bar{\Delta}_{\rm PG}({\bf k}_{\rm F})$ exhibits the largest value around the antinodes, however, the actual minimum of $\bar{\Delta}_{\rm PG}({\bf k}_{\rm F})\sim 0$ does not appear around the nodes, but locates exactly at the tips of the Fermi arcs, which is well consistent with the experimental observations \cite{Comin16,Wu11,Chang12,Ghiringhelli12,Comin14,Neto14,Fujita14,Campi15,Comin15a,Peng16,Hinton16,Hamidian16}. This phenomenological Hamiltonian (\ref{CDW-model}) consists of two parts, the CDW part with the momentum dependence of the CDW gap $\bar{\Delta}_{\rm PG}({\bf k})$, and the SC part with the momentum dependence of the SC gap $\bar{\Delta}_{\rm s}({\bf k})$. In this case, two basic low-energy excitations for the SC quasiparticle and CDW quasiparticle, respectively, should emerge as the propagating modes, with the scattering of the SC quasiparticles that mainly are responsible for superconductivity, while the scattering of the CDW quasiparticles dominates the CDW fluctuation. It should be emphasized that this type Hamiltonian (\ref{CDW-model}) has been usually employed to phenomenologically discuss the physical behavior of the CDW order and of its interplay with superconductivity in cuprate superconductors \cite{Sachdev13,Meier13,Harrison14,Atkinson15}.

In the framework of the equation of motion, the time-Fourier transform of the single-particle diagonal and off-diagonal Green's functions $G({\bf k}, \omega)$ and $\Im^{\dagger}({\bf k},\omega)$ of the Hamiltonian (\ref{CDW-model}) satisfies the following equations \cite{Mahan81},
\begin{subequations}\label{equations-of-motion}
\begin{eqnarray}
&~&(\omega-\varepsilon_{\bf k})G({\bf k},\omega)+\bar{\Delta}_{\rm s}({\bf k})\Im^{\dagger}({\bf k},\omega)\nonumber\\
&+& \bar{\Delta}_{\rm PG}({\bf k}) <T_{\tau}C_{{\bf k}+{\bf Q}\sigma}(\tau)C^{\dagger}_{{\bf k}\sigma}(\tau')>_{\omega} = 1, ~~~~~ \label{diagonal-equation}\\
&~&(\omega+\varepsilon_{\bf k})\Im^{\dagger}({\bf k},\omega)+\bar{\Delta}_{\rm s}({\bf k})G({\bf k},\omega)\nonumber\\
&+& \bar{\Delta}_{\rm PG}({\bf k}) <T_{\tau}C^{\dagger}_{{\bf k}+{\bf Q}\uparrow}(\tau)C^{\dagger}_{-{\bf k}\downarrow}(\tau')>_{\omega} = 0 . ~~~~~ \label{off-diagonal-equation}
\end{eqnarray}
\end{subequations}
However, it is clear from these equations that for obtaining the single-particle diagonal and off-diagonal Green's functions $G({\bf k},\omega)$ and $\Im^{\dagger}({\bf k},\omega)$, we need to introduce another two Green's functions,
\begin{subequations}\label{CDW-PDW-Green-functions}
\begin{eqnarray}
G_{\rm CDW}({\bf k},\tau-\tau')&=&-<T_{\tau}C_{{\bf k}+{\bf Q}\sigma}(\tau)C^{\dagger}_{{\bf k}\sigma}(\tau')>, \label{CDW-Green-function} ~~~~\\
\Im^{\dagger}_{\rm PDW}({\bf k},\tau-\tau')&=&<T_{\tau}C^{\dagger}_{{\bf k}+{\bf Q}\uparrow}(\tau)C^{\dagger}_{-{\bf k}\downarrow}(\tau')>, ~~~~ \label{PDW-Green-function}
\end{eqnarray}
\end{subequations}
which therefore describe the CDW and PDW states, respectively. After a straightforward calculation, we find that the time-Fourier transform of the CDW and PDW Green's functions $G_{\rm CDW}({\bf k},\omega)$ and $\Im^{\dagger}_{\rm PDW}({\bf k},\omega)$ satisfies the following equations \cite{Mahan81},
\begin{subequations}\label{CDW-PDW-equations-of-motion}
\begin{eqnarray}
G_{\rm CDW}({\bf k},\omega)&=&{\bar{\Delta}_{\rm PG}({\bf k})\over \omega+\varepsilon_{0\bf k}}G({\bf k},\omega),~~~~~ \label{CDW-equations-of-motion}\\
\Im^{\dagger}_{\rm PDW}({\bf k},\omega)&=&-{\bar{\Delta}_{\rm PG}({\bf k})\over \omega-\varepsilon_{0\bf k}}\Im^{\dagger}({\bf k},\omega). ~~~~~ \label{PDW-equations-of-motion}
\end{eqnarray}
\end{subequations}
Substituting these CDW and PDW Green's functions in Eq. (\ref{CDW-PDW-equations-of-motion}) into Eq. (\ref{equations-of-motion}), the single-particle diagonal and off-diagonal Green's functions $G({\bf k},\omega)$ and $\Im^{\dagger}({\bf k},\omega)$ of the Hamiltonian (\ref{CDW-model}) are obtained explicitly, and then these obtained single-particle diagonal and off-diagonal Green's functions $G({\bf k},\omega)$ and $\Im^{\dagger}({\bf k},\omega)$ are the exactly same as quoted in Eq. (\ref{EGF}) and the related the electron self-energy in Eq. (\ref{PG}).

From the above discussions, we therefore confirm that the single-particle diagonal and off-diagonal Green's functions $G({\bf k},\omega)$ and $\Im^{\dagger}({\bf k},\omega)$ in Eq. (\ref{EGF}) and the related the electron self-energy in Eq. (\ref{PG}) obtained based on the kinetic-energy-driven superconductivity can be reproduced exactly from the phenomenological Hamiltonian (\ref{CDW-model}) in terms of the CDW and PDW Green's functions $G_{\rm CDW}({\bf k},\omega)$ and $\Im^{\dagger}_{\rm PDW}({\bf k},\omega)$, which therefore also reveal clearly the secrets of the PDW order: (i) As in the case of the CDW order, the appearance of the PDW Green's function $\Im^{\dagger}_{\rm PDW}({\bf k}, \omega)$ in the calculation of the single-particle diagonal and off-diagonal Green's functions $G({\bf k},\omega)$ and $\Im^{\dagger}({\bf k}, \omega)$ also means the existence of the PDW order as shown in Fig. \ref{spectrum-maps}a (dashed-red lines) and its coexistence with the CDW order and superconductivity in the SC-state. However, since the Hamiltonian (\ref{CDW-model}) describes obviously the interplay between the CDW order and superconductivity only, the automatical appearance of the PDW order is therefore generated by the interplay of the CDW order with superconductivity, and can be thought to be the subsidiary order parameter; (ii) However, in a clear contrast to the case of the CDW order, the appearance of the PDW order does not need to introduce an obvious PDW order parameter in the Hamiltonian (\ref{CDW-model}), in other words, the appearance of the PDW order does not produce an ordered gap in the quasiparticle excitation spectrum, and in this sense, this PDW order is a novel hidden order.

In the normal-state, the SC gap $\bar{\Delta}_{\rm s}({\bf k})=0$, and then the phenomenological Hamiltonian (\ref{CDW-model}) is reduced as,
\begin{eqnarray}\label{NCDW-model}
H&=&\sum_{{\bf k}\sigma}\varepsilon_{\bf k}C^{\dagger}_{{\bf k}\sigma}C_{{\bf k}\sigma}-\sum_{{\bf k}\sigma}\varepsilon_{0{\bf k}}C^{\dagger}_{{\bf k} +{\bf Q}\sigma}C_{{\bf k}+{\bf Q}\sigma} \nonumber\\
&+&\sum_{{\bf k}\sigma}\bar{\Delta}_{\rm PG}({\bf k})(C^{\dagger}_{{\bf k}+{\bf Q}\sigma}C_{{\bf k}\sigma}+C^{\dagger}_{{\bf k}\sigma} C_{{\bf k}+ {\bf Q}\sigma}), ~~~~~~~~~
\end{eqnarray}
where the Hamiltonian (\ref{NCDW-model}) consists of the CDW part only with the momentum dependence of the CDW gap $\bar{\Delta}_{\rm PG}({\bf k})$. In this case, only one basic low-energy excitation for the CDW quasiparticle emerges as the propagating mode. In particular, the single-particle Green's function $G({\bf k},\omega)$ of the Hamiltonian (\ref{NCDW-model}) is evaluated in terms of the CDW Green's function $G_{\rm CDW}({\bf k}, \omega)$ only, and then the obtained single-particle Green's function is also the exactly same as the single-particle Green's function obtained based on the kinetic-energy-driven superconductivity and the related the electron self-energy in the normal-state \cite{Feng16,Zhao16,Feng15a}. In this normal-state case, the interplay of the CDW order with superconductivity disappears, leading to that the novel hidden PDW order is therefore absent in the normal-state.

Finally, we emphasize that since the CDW gap in the Hamiltonians (\ref{CDW-model}) and (\ref{NCDW-model}) has been identified as the pseudogap $\bar{\Delta}_{\rm PG}$, this is also why the CDW order exists within the pseudogap phase, appearing below a temperature of the order of $T^{*}$ well above $T_{\rm c}$ in the underdoped regime, and coexists with the hidden PDW order and superconductivity below $T_{\rm c}$ \cite{Comin16,Wu11,Chang12,Ghiringhelli12,Comin14,Neto14,Fujita14,Campi15,Comin15a,Peng16,Hinton16,Hamidian16,Badoux16}.

In conclusion, within the framework of the kinetic-energy-driven superconductivity, we have studied the physical origin of the PDW order in cuprate superconductors and of its connection with the interplay of the CDW order with superconductivity by taking into account the pseudogap effect. Our results show that the PDW order is generated by the interplay of the CDW order with superconductivity, and therefore automatically emerges as a subsidiary order parameter in the SC-state. However, the onset of the PDW order does not produce an ordered gap in the quasiparticle excitation spectrum, and in this sense, this PDW order is a novel hidden order. The theory also predicts that this novel hidden PDW order appeared automatically as a subsidiary order parameter in the SC-state is absent from the normal-state, which should be verified by future experiments.

\section*{Acknowledgements}

The authors would like to thank Professor Yongjun Wang for helpful discussions. This work was supported by the National Key Research and Development Program of China under Grant No. 2016YFA0300304, and the National Natural Science Foundation of China (NSFC) under Grant Nos. 11574032 and 11734002.


\begin{thebibliography}{00}

\bibitem{Hufner08} See, {\it e.g.}, H\"ufner, S. {\it et al}., Rep. Prog. Phys. {\bf 71}, 062501 (2008).

\bibitem{Basov05} See, {\it e.g.}, Basov, D. N. and Timusk, T., Rev. Mod. Phys. {\bf 77}, 721 (2005).

\bibitem{Timusk99} See, {\it e.g.}, Timusk, T. and Statt, B., Rep. Prog. Phys. {\bf 62}, 61 (1999).

\bibitem{Norman98} Norman, M. R. {\it et al}., Nature {\bf 392}, 157 (1998).

\bibitem{Yoshida06} Yoshida, T. {\it et al}., Phys. Rev. B {\bf 74}, 224510 (2006).

\bibitem{Kanigel07} Kanigel, A. {\it et al}., Phys. Rev. Lett. {\bf 99}, 157001 (2007).

\bibitem{Yoshida09} Yoshida, T. {\it et al}., Phys. Rev. Lett. {\bf 103}, 037004 (2009).

\bibitem{Yang08} Yang, H.-B. {\it et al}., Nature {\bf 456}, 77 (2008).

\bibitem{Meng09} Meng, J. {\it et al}., Nature {\bf 462}, 335 (2009).

\bibitem{Yang11} Yang, H.-B. {\it et al}., Phys. Rev. Lett. {\bf 107}, 047003 (2011).

\bibitem{Comin16} See, e.g., the review, Comin, R. and Damascelli, A., Annu. Rev. Condens. Matter Phys. {\bf 7}, 369 (2016).

\bibitem{Wu11} Wu, T. {\it et al}., Nature {\bf 477}, 191 (2011).

\bibitem{Chang12} Chang, J. {\it et al}., Nat. Phys. {\bf 8}, 871 (2012).

\bibitem{Ghiringhelli12} Ghiringhelli, G. {\it et al}., Science {\bf 337}, 821 (2012).

\bibitem{Comin14} Comin, R. {\it et al}., Science {\bf 343}, 390 (2014).

\bibitem{Neto14} da Silva Neto, Eduardo H. {\it et al}., Science {\bf 343}, 393 (2014).

\bibitem{Fujita14} Fujita, K. {\it et al}., Science {\bf 344}, 612 (2014).

\bibitem{Campi15} Campi, G. {\it et al}., Nature {\bf 525}, 359 (2015).

\bibitem{Comin15a} Comin, R. {\it et al}., Nat. Mater. {\bf 14}, 796 (2015).

\bibitem{Peng16} Peng, Y. Y. {\it et al}., Phys. Rev. B {\bf 94}, 184511 (2016).

\bibitem{Hinton16} Hinton, J. P. {\it et al}., Sci. Rep. {\bf 6}, 23610 (2016).

\bibitem{Hamidian16} Hamidian, M. H. {\it et al}., Nature {\bf 532}, 343 (2016).

\bibitem{Badoux16} Badoux, S. {\it et al}., Nature {\bf 531}, 210 (2016).

\bibitem{Sachdev13} Sachdev, S. and Placa, R. L., Phys. Rev. Lett. {\bf 111}, 027202 (2013).

\bibitem{Meier13} Meier, H. {\it et al}., Phys. Rev. B {\bf 88}, 020506(R) (2013).

\bibitem{Harrison14} Harrison, N. and Sebastian, S. E., New J. Phys. {\bf 16}, 063025 (2014).

\bibitem{Atkinson15} Atkinson, W. A. {\it et al}., New J. Phys. {\bf 17}, 013025 (2015).

\bibitem {Feng16} Feng, S. {\it et al}., Phil. Mag. {\bf 96}, 1245 (2016); Mou, Y. {\it et al}., Phil. Mag. {\bf 97}, 3361 (2017); Zhao, H. {\it et al}., J. Supercond. Nov. Magn. {\bf 31}, 683 (2018).

\bibitem {Zhao16} Zhao, H. {\it et al}., J. Supercond. Nov. Magn. {\bf 29}, 3027 (2016).

\bibitem {Gao18} Gao, D. {\it et al}., Physica C {\bf 551}, 72 (2018).

\bibitem{Lee14} Lee, P. A., Phys. Rev. X {\bf 4}, 031017 (2014).

\bibitem{Fradkin15} See, {\it e.g.}, Fradkin, E. {\it et al}., Rev. Mod. Phys. {\bf 87}, 457 (2015).

\bibitem{Wang15} Wang, Y. {\it et al}., Phys. Rev. B {\bf 91}, 115103 (2015).

\bibitem {Feng0306} Feng, S., Phys. Rev. B {\bf 68}, 184501 (2003); Feng, S. {\it et al}., Physica C {\bf 436}, 14 (2006); Feng, S. {\it et al}., Phys. Rev. B. {\bf 85}, 054509 (2012).

\bibitem {Feng15} See, {\it e.g.}, the review, Feng, S. {\it et al}., Int. J. Mod. Phys. B {\bf 29}, 1530009 (2015).

\bibitem{Feng15a} Feng, S. {\it et al}., Physica C {\bf 517}, 5 (2015).

\bibitem{Gao18a} Gao, D. {\it et al}., J. Low Temp. Phys. {\bf 192}, 19 (2018).

\bibitem{Damascelli03} See, {\it e.g}., the review, Damascelli, A. {\it et al}., Rev. Mod. Phys. {\bf 75}, 473 (2003).

\bibitem{Campuzano04} See, {\it e.g.}, the review, Campuzano, J. C. {\it et al}., In: Bennemann, K. H. and Ketterson, J. B. (eds.) Physics of Superconductors, vol. II, p. 167. Springer, Berlin Heidelberg New York (2004).

\bibitem{Zhou07} See, {\it e.g.}, the review, Zhou, X. J. {\it et al}., In: Schrieffer, J. R. (eds.) Handbook of High-Temperature Superconductivity: Theory and Experiment, p. 87. Springer, New York (2007).

\bibitem {Zhao17} Zhao, H. {\it et al}., Physica C {\bf 534}, 1 (2017).

\bibitem{Mahan81} See, {\it e.g.}, Mahan, G. D., Many-Particle Physics. Plenum Press, New York (1981).

\end{thebibliography}
\end{document}